\documentclass[runningheads]{llncs}
\usepackage[T1]{fontenc}
\usepackage{graphicx}
\usepackage{hyperref}
\usepackage{fancyref}
\usepackage{color}
\usepackage{placeins}
\usepackage{booktabs}

\usepackage{amsmath}
\usepackage{enumitem}

\begin{document}

\title{Augment to Augment: Diverse Augmentations Enable Competitive Ultra-Low-Field MRI Enhancement}
\titlerunning{Augment to Augment}

\author{Felix F Zimmermann\orcidID{0000-0002-0862-8973}}
\authorrunning{F.F. Zimmermann}

\institute{Physikalisch-Technische Bundesanstalt (PTB), Braunschweig and Berlin, Germany
\email{felix.zimmermann@ptb.de}}

\maketitle 

\begin{abstract}
Ultra-low-field (ULF) MRI promises broader accessibility but suffers from low
signal-to-noise ratio (SNR), reduced spatial resolution, and contrasts that
deviate from high-field standards. Image-to-image translation can map ULF
images to a high-field appearance, yet efficacy is limited by scarce paired
training data. Working within the ULF-EnC challenge constraints (50 paired
3D volumes; no external data), we study how task-adapted data augmentations
impact a standard deep model for ULF image enhancement. We show that strong,
diverse augmentations, including auxiliary tasks on high-field data, substantially
improve fidelity. Our submission ranked third by brain-masked SSIM on the
public validation leaderboard and fourth by the official score on the final test leaderboard.

Code is available at
\href{https://github.com/fzimmermann89/low-field-enhancement}{github.com/fzimmermann89/low-field-enhancement}.

\keywords{ULF-EnC \and ultra-low-field MRI \and image enhancement}
\end{abstract}

\section{Introduction}
Magnetic Resonance Imaging (MRI) is a cornerstone of medical diagnostics, with clinical practice dominated by systems operating at high field strengths of \(1.5\,\text{T}\) and \(3\,\text{T}\). These systems rely on sophisticated superconducting magnets, rendering them large, costly, and complex to site. In contrast, ultra-low-field (ULF) MRI, which operates at field strengths below \(100\,\text{mT}\), has emerged as a promising alternative\cite{arnold2023low}. The potential advantages of ULF MRI include reduced cost, enhanced safety, and fewer image artifacts from metallic implants or foreign bodies.

However, these benefits are counterbalanced by a fundamental physical limitation: the signal amplitude is proportional to the magnetic field strength, leading to an inherently low signal-to-noise ratio (SNR) in ULF images\cite{physics,arnold2023low}. This low SNR manifests as reduced achievable spatial resolution and increased image noise. Furthermore, the field-dependent nature of tissue relaxation mechanisms results in image contrasts that differ significantly from established clinical standards, posing an interpretative challenge for radiologists.

Efforts to overcome the limitations of ULF MRI are advancing on two primary fronts. The first involves innovations in hardware and data acquisition, including the design of novel acquisition schemes and improvements in signal amplification and noise rejection. The second front, which is the focus of this work, concentrates on image reconstruction and post-processing\cite{schote2022physics,ayde2025mri,arnold2023low,dayarathna2024ultra,dayarathna2024review}. This approach leverages computational imaging techniques and deep learning to mitigate the effects of low SNR and bridge the contrast gap to high-field standards.

Deep learning for ULF enhancement proceeds via two broad strategies. Physics-informed learned reconstruction integrates networks into the image formation from k-space with data consistency \cite{dl4,koonjoo2021boosting,shimron}, an approach that is state-of-the-art for accelerated high-field MRI
\cite{sriram_end--end_2020,aggarwal_modl_2019,nosense,promptmr}. This approach requires access to raw data, which is commonly unavailable on commercial systems. Alternatively, post-processing
performs image-to-image translation on conventional reconstructions \cite{islam2023improving,dl1}, which we adopt.

We operate within the ULF-EnC challenge\cite{ulf} setting: 50 paired 3D brain volumes
acquired at \(64\,\text{mT}\) (Hyperfine Swoop) and \(3\,\text{T}\) (Siemens Biograph mMR) for
T1-weighted, T2-weighted, and FLAIR, co-registered for paired training and evaluation. Our submission ranked third by brain-masked SSIM on the public validation leaderboard  and ultimately placed fourth on the final test leaderboard.

\section{Methods}
\begin{figure}[bt!]
    \centering
    \includegraphics[width=\linewidth]{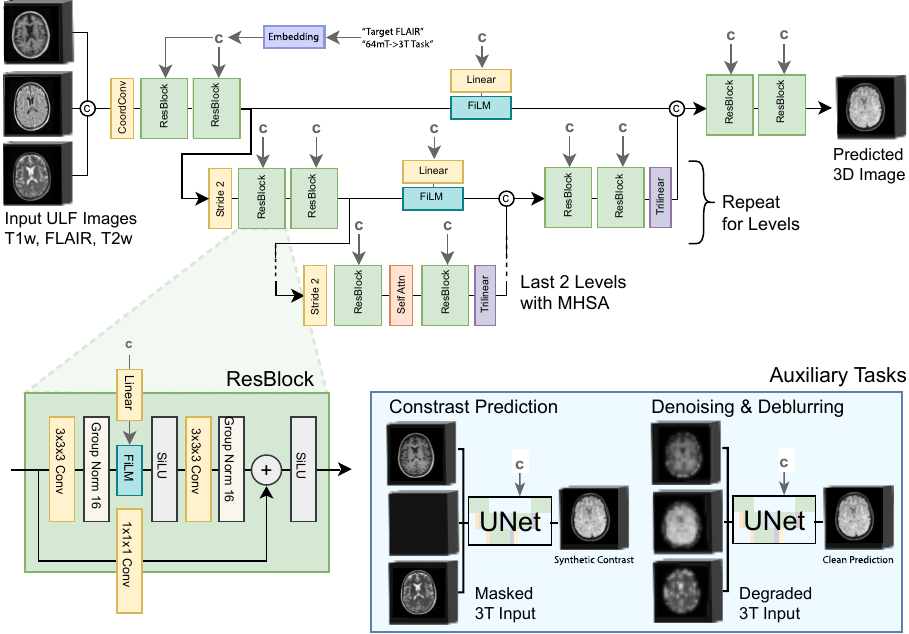}
    \caption{Schematic overview of the proposed method. A single 3D U-Net is trained on three tasks. The primary task (top) is to translate the three low-field contrasts into a target high-field contrast, specified via FiLM conditioning. Two auxiliary tasks augment the training by leveraging only the high-field data: (1) contrast synthesis and (2) image restoration (denoising/deblurring).}
    \label{fig:method}
\end{figure}

Our method for low- to high-field MRI translation is based on three key components: a multi-task network architecture, a hybrid training objective, and a strong data augmentation strategy, as illustrated in Figure \ref{fig:method}.

The full source code and trained weights as submitted to the challenge are available at \url{https://github.com/fzimmermann89/low-field-enhancement}.

The model was trained for 500 epochs using AdamW \cite{adamw} with a cosine learning rate schedule, taking approximately 24 hours on four NVIDIA H100 GPUs.

\subsection{Network Architecture}
A single multi-contrast 3D U-Net ingests T1-weighted, T2-weighted, and FLAIR volumes as channels. The target contrast and task is provided via an embedding and used for FiLM conditioning ($c$ in Fig. \ref{fig:method}). The architecture follows a standard 3D U-Net \cite{ronneberger_u-net_2015} as implemented in MRpro \cite{mrpro} with four resolution levels with 64$\rightarrow$128$\rightarrow$192$\rightarrow$256 channels. Each encoder/decoder level comprises two ResNet blocks with two convolutions each, SiLU activation, Group Normalization \cite{groupnorm}, and FiLM \cite{film} in the ResNet blocks and the skip connections. Multi-head self-attention is inserted at the two lowest resolutions for long-range context. We use strided convolutions for downsampling and trilinear interpolation for upsampling. A coordinate grid is concatenated to the inputs of the first convolution to provide explicit spatial context \cite{coordconv}. The model has approximately 90\,M parameters.

\subsection{Training Objective}
The training objective combines a supervised reconstruction loss with an adversarial loss \cite{goodfellow,dayarathna2025mccad,dayarathna2024ultra}. The reconstruction loss is a weighted sum of a pixel-wise L1 loss and a 3D structural similarity (SSIM) loss (rectangular 11x11x11 window), with weights 0.2 and 0.8, respectively. To mitigate the blurring often associated with pixel-wise losses, we incorporate an adversarial loss with weight 0.2. This is implemented using a conditional PatchGAN \cite{isola} discriminator (4 Levels of stride 2, 4x4x4 kernel convolutions, 32$\rightarrow$64$\rightarrow$128$\rightarrow$256 features, LeakyReLU activations, GroupNorm) within a hinge loss formulation \cite{hinge}. Conditioning on the target sequence is achieved by embedding the target into an 8-dimensional vector, broadcasting, and concatenating in the channel dimension. The discriminator is regularized with an R1 gradient penalty \cite{mescheder2018training} every second step.

\subsection{Auxiliary Tasks}
In addition to the primary task of low-to-high field translation, we train the network on two auxiliary tasks that leverage only the provided high-field data: high-field contrast synthesis and high-field image restoration\cite{noise2noise}. To accommodate these, the network input is designed to accept both low-field and high-field volumes in separate channels. For the main translation task, the high-field input channels are zeroed; for the auxiliary tasks, the low-field input channels are zeroed. For contrast synthesis, the network must learn to recreate a target contrast from the other two available high-field contrasts; this is achieved by zeroing out the target contrast channel in the input\cite{zimmermann2023semi}. For image restoration, the network is trained as a denoiser and deblurrer on randomly degraded high-field images.

\subsection{Data Augmentations}
Given the limited training data and the challenge's prohibition on external data, a robust augmentation strategy is essential\cite{augment,isensee2021nnu}. Our approach integrates auxiliary training tasks with a diverse set of standard data augmentations.

\paragraph{Geometric Augmentations}
We apply a set of random geometric augmentations consistently across the input contrasts and the target volume. These include standard affine transformations (rotations, shifts, shearing), left-right flips, and moderate 3D non-rigid transformations.

\paragraph{Intensity augmentations}
To improve robustness against variations in input contrast, a random monotonic intensity mapping is applied to each input volume in 20\% of training samples. For each contrast, four support values are randomly sampled from the range \([0, 1]\). These values are sorted and then assigned as the output intensities corresponding to the fixed input intensities of 0.2, 0.4, 0.6, and 0.8. A complete, continuous mapping is generated via linear interpolation between these support points, creating artificial variations in the input image contrast.

\paragraph{Input degradations}
To further increase robustness, the input volumes are additionally degraded in 20\% of training samples\cite{zhang2017beyond}. This involves adding Gaussian noise with a small, randomly selected standard deviation to each input contrast independently. Furthermore, we apply anisotropic blurring with a randomized kernel strength for each spatial dimension.

\section{Results}
\begin{figure}[t!]
    \centering
    \includegraphics[width=0.75\linewidth]{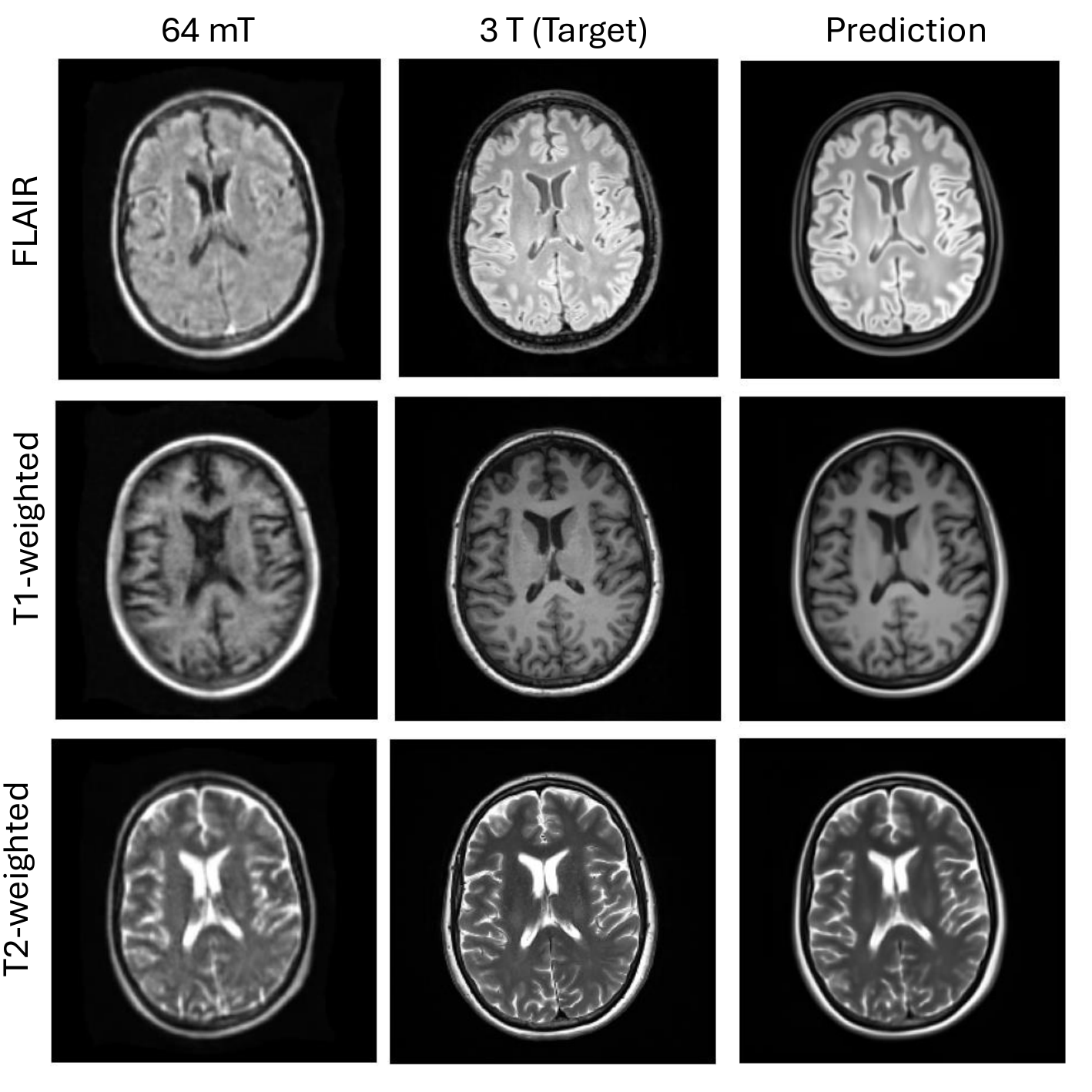}
    \caption{Exemplary result of the central axial slice of a held-out training sample. Given FLAIR (top), T1-weighted (center), and T2-weighted (bottom) input volumes (left), our trained network predicts 3\,T-like image volumes (right), closely matching the ground truth targets (center).}
    \label{fig:example}
\end{figure}

Two training subjects were held out for internal validation. Exemplary central axial slices from one validation subject are shown in \fref{fig:example}. In all three contrasts, the predicted enhanced images are visually close to the target 3\,T images. The SSIM values for this volume are 0.79/0.82/0.83 for the FLAIR/T1-/T2-weighted contrasts. Official challenge results on the official validation and test sets are reported in \fref{tab:val}.
The score used by the challenge, a combination of different metrics
$$
\text{Score}=0.7\,\text{SSIM} + 0.1\,\frac{\text{PSNR}}{32.0} + 0.1\,(1 - \text{MAE}) + 0.1\,(1 - \text{NMSE})
$$
was 0.779, placing our submission 4th out of 26 submissions from 18 teams.

\begin{table}
    \caption{Results of our proposed method on the validation and test set as reported by the official leaderboards.}
    \label{tab:val}
    \centering

\begin{tabular}{l@{\extracolsep{3em}}rrr}
\hline
\hline
& \multicolumn{2}{c}{Validation} & \multicolumn{1}{c}{Test} \\
\cmidrule(lr){2-3} \cmidrule(lr){4-4}
Metric & Value & Value Masked & Value Masked \\
\hline
SSIM   & 0.823  & 0.715        & 0.714         \\
PSNR   & 23.02  & 30.01        & 29.84       \\
MAE    & 0.032  & 0.069        & 0.070         \\
NMSE   & 0.067  & 0.153        & 0.066         \\
\hline
\hline
\end{tabular}
\end{table}

\subsection{Ablation Study}
We investigate the influence of different components of our method. The ablations were chosen to isolate the impact of our augmentation strategy (a, b, c), the choice of loss function (d), and the 3D network architecture (e):
\begin{enumerate}[label=\alph*)]
    \item \textit{No Augmentations}: All augmentations and auxiliary tasks removed.
    \item \textit{Only Affine Augmentations}: Only standard random affine transformation and flips. 
    \item \textit{No Auxiliary Tasks}: All data augmentations, but no auxiliary tasks.
    \item \textit{No Adversarial Loss}: Training with only the L1+SSIM reconstruction loss.
    \item \textit{2D U-Net}: A 2D U-Net trained on axial slices instead of the proposed 3D U-Net.
\end{enumerate}
The results on the internal validation samples, shown in \fref{tab:ablation}, highlight the substantial performance benefit of our diverse augmentation strategy. The 3D approach and the adversarial loss also prove to be critical components.

\begin{table}[]
\caption{Ablation study on held-out training subjects (internal validation).}
    \label{tab:ablation}
    \centering
\begin{tabular}{l@{\extracolsep{3em}}r}
\hline
\hline
Ablation & SSIM  \\
\hline
a) No Augmentation  & 0.68 \\
b) Only Affine Augmentations   & 0.76      \\
c) No Auxiliary Tasks   &  0.79   \\
d) No Adversarial Loss& 0.80\\  
e) 2D U-Net&     0.72  \\
\hline
\textbf{Proposed} & \textbf{0.82}\\ 
\hline
\hline
\end{tabular}
\end{table}
\FloatBarrier 
\section{Discussion}
We presented a deep learning method for ULF MRI enhancement that combines a
contrast-conditioned 3D U-Net, a hybrid reconstruction–adversarial objective,
and a diverse augmentation strategy. The ablation study confirmed the importance of our multi-faceted augmentation approach for achieving high performance with limited data. Challenge results independently demonstrate the competitiveness of this strategy. 

Our design choices were guided by observations of the training data. We noted that for a significant number of training samples, the co-registration between the \(3\,\text{T}\) and \(64\,\text{mT}\) volumes appeared suboptimal. This motivated our decision to use a 3D, multi-contrast approach, where all three source contrasts are used to predict a single target volume, allowing the network to leverage robust anatomical information from one contrast to aid the reconstruction of another.

A general limitation of learned image enhancement is the difficulty in assessing model uncertainty. The enhanced images may appear sharp and convincing even when generated from poor-quality inputs, potentially masking uncertainty or hallucinating anatomical details. This ill-posed problem is a critical barrier to clinical translation. 
To address this, we briefly investigated a diffusion-based model capable of sampling from the posterior distribution of possible high-field images. While this approach effectively visualizes model uncertainty by generating multiple, distinct high-field reconstructions for a single input, its performance on standard metrics was not competitive within the challenge constraints. We hypothesize this can be addressed by using additional, unpaired public training data and plan to pursue this avenue further.

\begin{credits}
\subsubsection{\ackname} This work was supported in
part by the Metrology for Artificial Intelligence for Medicine (M4AIM) Project that is funded
by the German Federal Ministry for Economic Affairs and
Climate Action (BMWi) in the
framework of the QI-Digital initiative and in part by the
Deutsche Forschungsgemeinschaft (DFG, German Research
Foundation) under Grant 372486779 (SFB1340).
\subsubsection{\discintname}
The authors have no competing interests to declare that are relevant to the content of this article.
\end{credits}

\bibliographystyle{splncs04}
\bibliography{bib}

\end{document}